\documentstyle[sprocl,psfig]{article}
\def\be{\begin{equation}}
\def\ee{\end{equation}}
\def\bea{\begin{eqnarray}}
\def\eea{\end{eqnarray}}
\def\simlt{\stackrel{<}{{}_\sim}}
\def\simgt{\stackrel{>}{{}_\sim}}

\begin{document}
\pagestyle{empty}
\title{STATUS OF LOW ENERGY SUPERSYMMETRY\footnote{Based on invited talks
given at ``RADCOR98'', Barcelona, September 1998 and ``Beyond the Standard
Model'', DESY, Hamburg, September 1998}}
\author{Stefan Pokorski}
\address{Institute of Theoretical Physics, Warsaw University, 
Ho\.za 69, 00-681 Warsaw, Poland.}
\maketitle\abstracts{We review 1) constraints on low energy supersymmetry
from  the search for Higgs boson and from precision data, 2) dependence
of coupling unification on the superpartner spectrum, 3) naturalness and fine
tuning in the minimal and non- minimal scenarios.}

\section{Constraints from the search for  Higgs boson and from
precision data}
\vskip 0.2cm

The most appropriate starting point for reviewing the status of low 
energy supersymmetry is the status of the Standard Model itself. Its
success in describing all the available experimental data becomes more
and more pronouced, with all potential deviations disappearing with
the increasing precision of data. At present, the Standard Model is 
successfully tested at $1$ permille  accuracy up to the LEP2 and
TEVATRON energies.

One of the most important results following from the precision tests
of the Standard Model is the strong indirect indication for a light
Higgs boson. Although the sensitivity of electroweak observables to
the Higgs boson mass is only logarithmic, the precision of both data
and calculations is high enough for obtaining from the fits the upper
bound on the Higgs boson mass $M_h$ of about 250 GeV at $95\%$ C.L.
The best value of $M_h$ in the fits is in the region of the present direct
experimental lower bound, $M_h\simgt 90$ GeV.

One should stress that these results are obtained strictly in the Standard
Model. The best fitted value of $M_h$ can be changed if we admit new 
physics in the $\Delta \rho$ parameter. There is the well known
(see, for instance, \cite{ellis}) ``flat direction'' 
in $\chi^2$,  which correlates
almost linearly $\ln M_h$ with $(\Delta \rho)^{NEW}$ (previously known as $\ln
M_h--m_t$ correlation), as the two effects compensate each other in
the electroweak observables like $\rho$,~ $sin^2\Theta_{eff}$ and $M_W$.
In general, however, it is very difficult to find a self-consistent
extension of the Standard Model that would use this freedom.
\footnote{Note also that $M_h\simlt {\cal O}(500)$ GeV by unitarity and 
``triviality'' arguments. So  the corrections to $\Delta \rho$
must be just right (not too big), to explore only relatively small
variation of $\ln M_h$.}
 
The result for $M_h$ from the Standard Model fits to precision data
raises  strong hopes for experimental discovery of the Higgs boson in
a relatively near future. Secondly, it is in agreement with the most
robust prediction of supersymmetric extensions of the Standard Model,
which is the existence of a light Higgs boson. This prediction is
generic for low energy effective supersymmetric models \cite{savoy,
kane} and becomes particularly quantitative in the Minimal
Supersymmetric Standard Model (MSSM), defined by three assumptions:
a)~minimal particle content consistent with the observed particle spectrum
and with supersymmetry, ~b)~R-parity conservation,~c)~most general soft 
supersymmetry breaking terms consistent with the SM gauge group.
\footnote{There is often some confusion about the terminology ``MSSM''. 
We always understand the MSSM as an effective low energy model with 
parameters unconstrained by any further high scale model assumptions.}

In the MSSM the lightest Higgs boson mass is predicted \cite{higgs,my} (now at
two- loop level \cite{HEHO,htwol}) in terms of free parameters of the model:
\begin{equation}
M_h=M_h(M_Z,G_\mu,\alpha_{EM},m_t,\tan\beta,M_A, {\rm superpartner~ masses})
\label{1}
\end{equation}
In practise, only the third generation sfermions are important in 
eq.(\ref{1}) and $M_h$ depends logarithmically on their masses. It is
worth recalling the dependence of $M_h$ on $\tan\beta$ and $M_A$ 
(see, for instance \cite{my}): for fixed $\tan\beta$ and superpartner
masses, $M_h$ reaches its maximal value for $M_A\approx 250$ GeV and
for larger values of $M_A$ it stays then approximately constant. As a
function of $\tan\beta$, these maximal values rise with $\tan\beta$
and remain almost constant for $\tan\beta\simgt 4$, with $M_h\approx 130$
GeV for superpartner masses lighter than 1 TeV. There is also a rather
strong dependence on the left-right mixing in the stop sector, with
the $\tan\beta$ dependent upper bounds for $M_h$ reached for large
mixing \cite{htwol}.

Clearly, the upper bound on $M_h$ from precision fits in the Standard
Model is encouraging for supersymmetry. However, one can also ask how 
{\it constraining} for the MSSM is the present direct  experimental
lower limit $M_h>90$ GeV. \footnote{Strictly speaking, this limit is
valid only in the Standard Model. In the MSSM, in some small regions
of  parameter space, the limit is actually lower, but we ignore
this effect here.}  This question has been studied in
ref. \cite{chank1} and the reader is invited to consult it for
details. The main conclusion is that, indeed, the low $\tan\beta$
region (interesting in the context of the quasi-infrared fixed point
scenario) is strongly constrained. It can be realized in Nature only
if at least one stop is heavy, ${\cal O}(1)$ TeV, and with large
left-right mixing. This follows from the fact that for low $\tan\beta$
the tree level lightest Higgs boson mass is small and large radiative
corrections have to account for the experimental bound. The infrared
fixed point scenario with low $\tan\beta$ will be totally ruled out if
a Higgs boson is not found at LEP2 operating at 200 GeV. For
intermediate and large values of $\tan\beta$, those constraints, of course,
disappear.

Superpartner masses that appear in radiative corrections to the Higgs
boson mass also appear in the calculation of $\Delta\rho$ and related 
observables such as $\sin^2\theta_{eff}$, $M_W$ etc. 
It is well known (see, for instance, \cite{unpublished,pokwar}) 
that the main new contribution to
$\Delta\rho$ comes from the third generation left-handed sfermions.
The custodial $SU_V(2)$ breaking in other sectors of the MSSM is very
weak. Thus, the quality of description of precision electroweak data
in the MSSM depends on those superpartner masses.
Instead of attempting an overall fit, it is more instructive to focus
on very well measured $\sin^2\theta_{eff}$ and on soon very well
measured  $M_W$. Calculating e.g. $\sin^2\theta_{eff}$ in terms of
$M_Z$, $G_\mu$, $\alpha_{EM}$ and the superpartner masses and comparing
with the experimental value we expect  to get bounds on the
left-handed third generation sfermions. We said earlier that for low 
$\tan\beta$ strong lower bounds on the stop mass follow from the
experimental lower limit on $M_h$. For such low values of $\tan\beta$ 
the bound from precision data is somewhat weaker  but, contrary to the 
other bound,  it remains very significant for all
values of $\tan\beta$. The absolute lower bound on the left-handed
stop and slepton masses is obtained for intermediate and large
$\tan\beta$ since  the data can then acommodate larger correction to
$\Delta\rho$ due to a heavier Higgs boson -see the earlier discusion.
In Fig.1 we show \cite{unpublished} 
the dependence of $\sin^2\Theta_{eff}$ and of $M_W$ on
the  stop and slepton masses. We conclude that for $2\sigma$ precision
in these observables one needs $m_{\tilde q_L}>{\cal O}(400)$ GeV and 
$m_{\tilde l_L}>{\cal O}(150)$ GeV, with stronger bounds in low $\tan\beta$
region. Precision data (and the lower limit on $M_h$ for low
$\tan\beta$) indicate that at least some superpartners are well above
the electroweak scale!

\begin{figure}
\psfig{figure=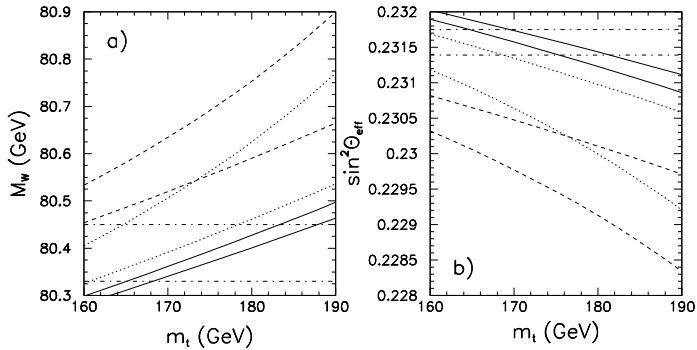,height=5.0cm}
\caption{{\bf a)} $W^\pm$ mass, and  {\bf b)} $\sin^2\theta^{eff}_{lept}$
as a function of the top quark mass, calculated for all but the heavier
stop and the heavier slepton superpartner masses
equal to 90 GeV. Top-down in {\bf a} and bottom-up curves in {\bf b}: dashed -
the heavier slepton (degenarate) masses at 90 GeV and the heavier stop masses
at 200 and 400 GeV, respectively; dotted -slepton masses at 150 GeV
and stop masses as before; solid - slepton masses at 250 GeV,
stop mass at 500 GeV and  at 1 TeV, respectively;
dashed-dotted horizontal curves-experimental 1$\sigma$ bands.}
\label{fig:barc1}
\end{figure}

On the other hand, the right-handed sfermions of the third generation
and all sfermions of the first two generations as well as the gaugino/
higgsino sectors are essentially unconstrained by the LEP precision
data. They decouple from those observables even if their masses are 
${\cal O}(M_Z)$.

Are there other observables that are more sensitive to the rest of the
spectrum, that is in which its decoupling in virtual effects is
slower?  Indeed, there are such examples. The decay $b\rightarrow
s\gamma$, the $K-{\bar K}$ and $B-{\bar B}$ mixing
are sensitive to the right-handed third generation
sfermions and to the higgsino-like chargino/neutralino. This is
because the relevant coupling is the top quark Yukawa coupling. 
For a review see \cite{misiak}. (Even
the gaugino and the first and the second generation sfermion
contribution decouples quite slowly in the $b\rightarrow s\gamma$ decay.)
Those processes have still good prospects to reveal indirect effects
of supersymmetry once the precision of data is improved.

The superpartner mass spectrum is the low energy window to 
the mechanism  of supersymmetry breaking and to the theory of
soft supersymmetry breaking terms. For instance, with the lower bound on 
the left-handed stop from the precision data and with the still open
possibility of a much lighter, say ${\cal O}(100)$ GeV, right-handed stop
one can envisage the case of a strongly split spectrum. This case is
discussed in ref.\cite{chank2} as an illustration  of the bottom-up approach
to the problem of supersymmetry breaking. It is shown that for low
$\tan\beta$ it needs, at the GUT scale, scalar masses much larger than the
gluino mass and the strong non-universality pattern, $m_Q^2:m_U^2:m^2_{H_2}=
1:2:3$. This is related to the fact that the hypothetical spectrum  considered
in this example departs from the well known  sum rules valid in the infrared
fixed point limit and with the scalar and gaugino masses of the same order
of magnitude. In such a scenario, the lighter chargino is generically also
light, ${\cal O}(100)$ GeV, and it can be gaugino- or higgsino-like.

\vskip 0.3cm

\section{Dependence of  coupling unification on the superpartner spectrum}
\vskip 0.2cm

The gauge coupling unification \cite{cos,zbyszek} within the MSSM has been 
widely publicized as the most important piece of indirect evidence for 
supersymmetry at accessible energies. The unification idea is predictive if 
physics at the GUT scale is described in terms of only two parameters: 
$\alpha_U$ and $M_U$. Then we can predict, for instance, $\alpha_s(M_Z)$ in 
terms of 
$\alpha_{EM}(M_Z)$ and $\sin^2\theta_W(M_Z)$. Here we mean the running 
coupling constants defined in the $\bar{MS}$ renormalization scheme in the 
SM which, we assume, is the correct renormalizable theory at the electroweak 
scale. The value of $\alpha_{EM}(M_Z)$ is obtained from the on-shell
$\alpha^{OS}_{EM}$=1/137.0359895(61) via the RG running in the SM, with .01\%
uncertainty due to the continous hadronic contribution to the photon 
propagator. The most precise value of $\sin^2\theta_W(M_Z)$ in the SM is at 
present obtained from its calculation in terms of $G_\mu,~M_Z,~\alpha_{EM}$ 
and the top quark mass (with some dependence on the Higgs boson mass). The 
unification prediction for $\alpha_s(M_Z)$ is obtained by using two loop RG 
equations in the MSSM, for the running from $M_Z$ up to the GUT scale defined 
by the crossing of the electroweak couplings. For the two--loop consistency 
(and precision), one must include the supersymmetric threshold corrections in 
the leading logarithmic approximation. (For a spectrum that is above the 
present experimental lower limits on the superpartner masses, the finite 
threshold effects ${\cal O}(\frac{M_Z}{m_{SUSY}})$ are already small enough 
to be neglected.\cite{zbyszek}) In this approximation the dependence of 
$\alpha_s(M_Z)$ on the supersymmetric spectrum can to a very good 
approximation be described by a single parameter $T_{SUSY}$ \cite{tsusy}. 
We get
\begin{equation}
\alpha_s(M_Z)=f(G_\mu, M_Z,\alpha_{EM},m_t,M_h,T_{SUSY})
\end{equation}
where
\begin{equation}
T_{SUSY}=|\mu|\left(\frac{m_{\tilde W}}{m_{\tilde g}})\right)^{3/2}
\left(\frac{M_{\tilde l}}{M_{\tilde q}}\right)^{3/16}
\left(\frac{M_{A^0}}{|\mu|}\right)^{3/19}\left(\frac{m_{\tilde W}}
{|\mu|}\right)^{4/19}
\end{equation}
We observe that the effective scale $T_{SUSY}$ depends strongly on the values 
of $\mu$ and of the ratio $m_{\tilde W}$ to $m_{\tilde g}$ but very weakly on 
the values of the squark and slepton masses. It is also clear that the scale 
$T_{SUSY}$ can be much smaller than $M_Z$ even if all superpartner masses are 
heavier than the $Z$ boson. Only for a fully degenerate spectrum $T_{SUSY}$ 
is the common superpartner mass. Moreover we note that the supersymmetric 
threshold effects are absent for $T_{SUSY}=M_Z$. The unification prediction 
for the strong coupling constant as a function of $T_{SUSY}$ is shown in 
Fig.2a.  We see that the variation of $\alpha_s(M_Z)$ with $T_{SUSY}$ is 
substantial. The central experimnetal value $\alpha_s(M_Z)=0.118$
is obtained for $T_{SUSY}=1$ TeV. The values $T_{SUSY}=M_Z$ and $T_{SUSY}=10$ 
TeV change the prediction by $\delta\alpha\approx\mp 0.01$ which is $3\sigma$ 
away from the central value. It is interesting to see that the value as 
large as $T_{SUSY}=10$ TeV is equally acceptable (or unacceptable) as 
$T_{SUSY}=M_Z$. As we already mentioned, a given value of $T_{SUSY}$ not 
necessarily implies similar scale for the superpartner masses. It should 
be stressed that in models with universal gaugino masses at the GUT scale we 
have the relation
\begin{equation}
\frac{m_{\tilde W}}{m_{\tilde g}}\approx\frac{\alpha_2(M_Z)}{\alpha_3(M_Z)}
~and~~~~ T_{SUSY}\approx |\mu|\left(\frac{\alpha_2(M_Z)}{\alpha_3(M_Z)}
\right)^{3/2}\approx\frac{1}{7}|\mu|
\end{equation}
Moreover, radiative electroweak symmetry breaking correlates the $\mu$ 
parameter  with the soft parameters that determine 
the sfermion masses, and large $\mu$ implies squark masses of the same order of
magnitude (but not vice versa!). Of course, it is also conceivable to have 
large $T_{SUSY}$ with small $\mu$. This requires $M_2/M_3>>1$ 
\cite{roszkowski} and, 
therefore, a gauge dependent transmission to the visible sector of the
supersymmetry breaking mechanism.

The minimal unification may be too restrictive as it is generally expected 
that there are some GUT/stringy threshold correction or higher dimension 
operator effects. Therefore is also interesting to reverse the question and 
to study the convergence in the MSSM of all the three couplings in the 
bottom-up approach, starting with their experimental values at the scale $M_Z$.

\begin{figure}
\psfig{figure=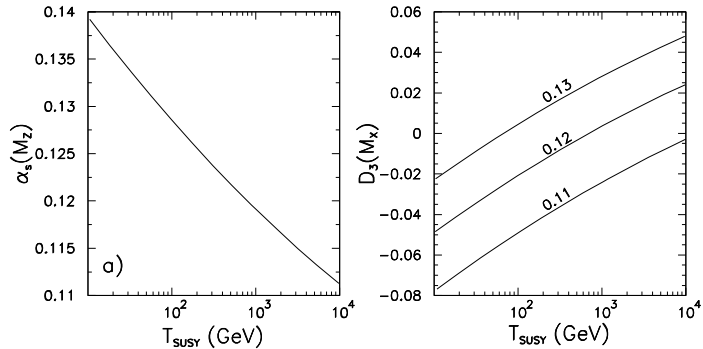,height=5.0cm}
\caption{ Unification prediction for  $\alpha_3(M_Z)$ ({\bf a}) and
$D_3$ defined by eq.(5) ({\bf b}), as a function of $T_{SUSY}$ ($m_t=175$ GeV,
$\tan\beta =2$)}
\label{fig:barc2}
\end{figure}

It is convenient to define the ``mismatch'' parameter \cite{vayonakis}
\begin{equation}
D_3={\alpha_3(M_{GUT})-\alpha_2(M_{GUT})\over\alpha_2(M_{GUT})}
\end{equation}
where $M_{GUT}$ is again defined by the crossing of the electroweak couplings.
The results for $D_3$ as a function of $T_{SUSY}$ are shown in Fig.2b
for three values of $\alpha_3(M_Z)$. We recall that the experimental value
is $\alpha_3(M_Z)=0.118\pm 0.003$.
We see that for $T_{SUSY}$ changing from $M_Z$ up to 10 TeV all the three 
couplings unify within 2\% accuracy! On the one hand, this is certainly an 
impressive success of the MSSM, but on the other hand we conclude that the 
gauge coupuling unification does not put any significant upper bounds on the 
superpartner spectrum.

Yukawa coupling unification is a much more model dependent issue. It strongly 
relies on GUT models and has no generic backing in string theory. Nevertheless,
it happens that in the bottom - up approach the $b$ and $\tau$ Yukawa 
couplings approximately unify at the same scale as the gauge couplings. On a 
more quantitative level, it is well known that exact $b-\tau$ Yukawa
coupling unification, at the  level of two-loop renormalization  group
equations   for  the  running   from the  GUT   scale down   to $M_Z$,
supplemented by three-loop QCD running down to the  scale $M_b$ of the
pole  mass  and finite two-loop   QCD  corrections at this  scale,  is
possible only for very small or very large values of $\tan\beta$. This
is due to the  fact  that renormalization of   the $b$-quark mass   by
strong interactions is too strong, and has to be partly compensated by
a large $t$-quark Yukawa coupling. This result is shown in Fig.~3a. We
compare there the running mass $m_b(M_Z)$ obtained by the running down
from    $M_{GUT}$, where we take $Y_b=Y_{\tau}$,     with the range of
$m_b(M_Z)$  obtained from   the    pole mass $M_b=(4.8\pm    0.2)$ GeV
\cite{BENEKE},  taking  into   account the above-mentioned  low-energy
corrections. These translate the range  of the pole mass: 4.6  $<M_b<$
5.0 GeV into   the following  range  of the   running mass $m_b(M_Z)$:
$2.72<m_b(M_Z)<3.16$~GeV.  To     remain   conservative,   we      use
$\alpha_s(M_Z)=0.115(0.121)$  to   obtain an  upper  (lower) limit  on
$m_b(M_Z)$. 

It is also well known \cite{HARASA,CAOLPOWA} that,  at least for large
values  of $\tan\beta$,  supersymmetric   finite one-loop  corrections
(neglected in Fig.~3a)   are  very important.  These corrections   are
usually not considered for intermediate values  of $\tan\beta$, but  
they  are also very  important there \cite{chank3} and make
$b-\tau$ unification viable in  much larger range of $\tan\beta$  than
generally believed (see also  \cite{MAPI}). 

One-loop diagrams with  bottom squark-gluino  and top  squark-chargino
loops   make  a contribution   to    the bottom-quark mass   which  is
proportional to $\tan\beta$ \cite{HARASA,CAOLPOWA}. We recall that, to
a good approximation, the one-loop correction to the bottom quark mass
is given by the expression: 
\begin{eqnarray}
{\Delta m_b\over m_b} \approx 
{\tan\beta\over4\pi}\mu\left[{8\over3}\alpha_s m_{\tilde g}
I(m^2_{\tilde g},M^2_{\tilde b_1},M^2_{\tilde b_2}) + 
Y_tA_tI(\mu^2,M^2_{\tilde t_1},M^2_{\tilde t_2})\right]\label{eqn:mbcorr}
\end{eqnarray}
where 
\begin{eqnarray}
I(a,b,c)=-{ab\log(a/b)+bc\log(b/c)+ca\log(c/a)\over(a-b)(b-c)(c-a)}
\nonumber
\end{eqnarray}
and  the   function $I(a,b,c)$  is   always positive and approximately
inversely proportional to its largest argument. This is the correction
to the running $m_b(M_Z)$. It is clear  from Fig.~3a that for $b-\tau$
unification in the intermediate $\tan\beta$ region  we need a negative
correction of order (15-20)\%  for  $3\simlt \tan\beta  \simlt 20$,
and about  a    10\%  correction for  $\tan\beta=30$. According to eq.
(\ref{eqn:mbcorr}),   such  corrections  require  $\mu<0$. 

We notice  that,   as   expected from    (\ref{eqn:mbcorr}),  $b-\tau$
unification       is   easier     for    $\tan\beta=30$    than    for
$\tan\beta\approx10$. In the latter case  it requires $A_t\simgt0$, in
order to obtain an enhancement   in (\ref{eqn:mbcorr}) or at least  to
avoid any cancellation between   the two terms  in (\ref{eqn:mbcorr}).
This is a strong  constraint on  the  parameter space. Since  $A_t$ is
given by \cite{CAOLPOWA1}:
\begin{equation}
A_t\approx (1-y)A_0-{\cal O}(1-2)M_{1/2}
\end{equation}
where $y=Y_t/Y^{FP}_t$ is the ratio of the top Yukawa coupling to its quasi-
infrared fixed point value, $b-\tau$ unification requires large positive
$A_0$  and not too   large  a $M_{\tilde   g}$ (i.e., $M_{1/2}$
for universal gaugino masses).   In
addition, the low-energy   value of $A_t$  is  then  always relatively
small and this implies a stronger upper  bound on $M_h$ (for a similar
conclusion, see   \cite{MAPI}).  We   see   in Fig.~3b     that,   for
$\tan\beta\simlt 10$, the  possibility of  exact $b-\tau$  unification
evaporates    quite  quickly, with    a   non-unification  window  for
$2\simlt\tan\beta\simlt8-10$, depending on  the  value of  $\alpha_s$.
However,  we also  see  that  supersymmetric one-loop corrections  are
large enough  to assure  unification within  10\% in almost  the whole
range of small and intermediate $\tan\beta$. 

For $\tan\beta >10$,  the  qualitative picture changes gradually.  The
overall factor  of  $\tan\beta$, on  the  one hand,  and  the need for
smaller corrections, on the other hand, lead  to the situation where a
partial   cancellation  of  the  two  terms   in (\ref{eqn:mbcorr}) is
necessary, or both   corrections  must be  suppressed  by sufficiently
heavy   squark masses.  Therefore,      $b-\tau$
unification for $\tan\beta=30$  typically requires a negative value of
$A_t$, and is  only marginally possible  for positive $A_t$, for heavy
enough squarks.  A similar but more extreme  situation occurs for very
large $\tan\beta$ values. It is
worth    recalling      that   the  second  term    in
(\ref{eqn:mbcorr}) is  typically at most of order  of (20-30)\% of the
first term \cite{CAOLPOWA}, due  to eq.(7). Thus, cancellation
of the  two  terms is  limited, and   for very  large $\tan\beta$  the
contribution of (\ref{eqn:mbcorr})   must  be  anyway   suppressed  by
requiring heavy squarks. This trend  is visible in Fig.~4a already for
$\tan\beta=30$.  The Higgs-boson mass is  not  constrained by $b-\tau$
unification,  since  $A_t$ can be  negative   and large.   

We  turn our     attention  now to  a  deeper    understanding of  the
$b\rightarrow   s\gamma$ constraint and   its interplay  with $b-\tau$
unification.  The  first  point   we would   like   to make   is  that
$b\rightarrow s \gamma$  decay is a  rigid  constraint in  the minimal
supergravity  model,  but is only  an   optional one  for the  general
low-energy   effective MSSM. Its   inclusion   depends on  the  strong
assumption that the stop-chargino-strange  quark  mixing angle is  the
same as the CKM element $V_{ts}$. This is the case only if squark mass
matrices are diagonal in the  super-KM basis,  which is realized,  for
instance,  in  the  minimal  supergravity   model. However,  for   the
right-handed up-squark sector such an  assumption is not imposed  upon
us by FCNC processes \cite{MASI}. Indeed,  aligning the squark flavour
basis with that of the quarks, the up-type squark right-handed flavour
off-diagonal mass squared matrix elements $(m^2_{\tilde U})^{13}_{RR}$
and $(m^2_{\tilde   U})^{23}_{RR}$  are unconstrained   by  other FCNC
processes. 

In   the  minimal  supergravity  model the  dominant  contributions to
$b\rightarrow s\gamma$  decay come from  the chargino-stop and charged
Higgs-boson/top-quark loops.  For  intermediate and large $\tan\beta$,
one  can estimate these   using the formulae of~\cite{BAGIbsg} in  the
approximation of  no mixing between  the gaugino  and higgsinos, i.e.,
for $M_W\ll max(M_2, |\mu|)$. We get \cite{BOOLPO} 
\begin{eqnarray}
{\cal A}_W&\approx&{\cal A}_0^\gamma{3\over2}{m_t^2\over M_W^2}f^{(1)}
                   \left({m_t^2\over M_W^2}\right)\\
{\cal A}_{H^+}&\approx&{\cal A}_0^\gamma{1\over2}{m_t^2\over M_{H^+}^2}
                      f^{(2)}\left({m_t^2\over M_{H^+}^2}\right)\\
{\cal A}_{C}&\approx&-{\cal A}_0^\gamma\left\{\left({M_W\over M_2}\right)^2
                \left[\cos^2\theta_{\tilde t}
                 f^{(1)}\left({M_{\tilde t_2}^2\over M_2^2}\right)
                +\sin^2\theta_{\tilde t}
                 f^{(1)}\left({M_{\tilde t_1}^2\over M_2^2}\right)\right]
                \right.\nonumber\\
            && -\left({m_t\over2\mu}\right)^2
                \left[\sin^2\theta_{\tilde t}
                 f^{(1)}\left({M_{\tilde t_2}^2\over\mu^2}\right)
                +\cos^2\theta_{\tilde t}
                 f^{(1)}\left({M_{\tilde t_1}^2\over\mu^2}\right)\right]
                \label{eqn:susy}\\
&&-\left.{\tan\beta\over2}{m_t\over\mu}
                {m_tA_t\over M^2_{\tilde t_2} - M_{\tilde t_1}^2}
                \left[f^{(3)}\left({M_{\tilde t_2}^2\over\mu^2}\right)
                - f^{(3)}\left({M_{\tilde t_1}^2\over\mu^2}\right)\right]
                \right\}\nonumber
\end{eqnarray}
where $\tilde t_1(\tilde t_2)$ denotes the lighter (heavier) stop,
\begin{equation}
\cos^2\theta_{\tilde t}={1\over2}\left(1 + \sqrt{1-a^2}\right), \phantom{aaaaa}
a\equiv{2m_tA_t\over M^2_{\tilde t_2} - M_{\tilde t_1}^2}, \phantom{aaaaa}
{\cal A}_0^\gamma\equiv G_F\sqrt{\alpha/(2\pi)^3}~V^\star_{ts}V_{tb}
\label{alwaysnumber}
\end{equation}
and the functions  $f^{(k)}(x)$ given in \cite{BAGIbsg} are  negative.
The  contribution ${\cal A}_{C}$  is  effectively proportional to  the
stop mixing parameter $A_t$, and  the sign of ${\cal A}_{C}$  relative
to ${\cal A}_W$ and ${\cal A}_{H^+}$ is negative for $A_t\mu<0$. 

We  can  discuss  now the  interplay of   the $b-\tau$ unification and
$b\rightarrow  s\gamma$ constraints.  The  chargino-loop  contribution
(\ref{eqn:susy}) has to be small or positive, since the Standard Model
contribution and  the  charged  Higgs-boson exchange  (both  negative)
leave  little room for  additional constructive contributions.  Hence,
one  generically   needs  $A_t\mu<0$.   Since  $\mu<0$   for  $b-\tau$
unification, both constraints  together  require $A_t>0$.  This is  in
line with our  earlier results for  the proper  correction to the  $b$
mass for  $\tan\beta\simlt 10$, \footnote{This  does not constrain the
parameter space more than $b-\tau$ unification itself. Note also that,
if  we  do not insist  on  $b-\tau$  unification,  the $b\rightarrow s
\gamma$ constraint is  easily  satisfied since $\mu>0$  is  possible.}
but  typically in conflict with such  corrections for larger values of
$\tan\beta$. In the latter case, both constraints can be satified only
at  the  expense  of  heavy  squarks (to   suppress   a positive $A_t$
correction to the $b$-quark  mass or  a  negative $A_t$  correction to
$b\rightarrow s \gamma$)  and a heavy   pseudoscalar $A^0$, as seen in 
Fig.4a.  

\begin{figure}
\psfig{figure=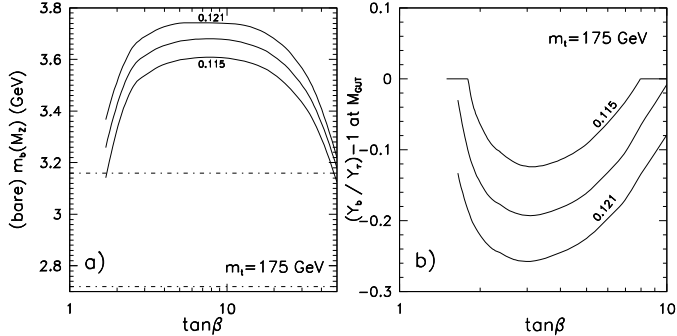,height=5.0cm}
\caption{{\bf a)} The running    mass $m_b(M_Z)$ obtained from strict
$b-\tau$ Yukawa coupling  unification at  $M_{GUT}=2\times10^{16}$ GeV
for different values  of $\alpha_s(M_Z)$, before inclusion of one-loop
supersymmetric corrections. {\bf b)} The minimal departure from $Y_b=Y_\tau$
at $M_{GUT}$ measured by the ratio $Y_b/Y_\tau -1$, which is necessary
for obtaining  the correct $b$  mass in the minimal supergravity model
with one-loop supersymmetric corrections included.}
\label{fig:barc3}
\end{figure}

\vskip 0.3cm

\section{Naturalness and fine tuning}
\vskip 0.2cm

The main theoretical motivation for the appearance of sparticles at the 
accessible energies is in order to alleviate the fine tuning required to
maintain the electroweak hierarchy, and sparticles become less effective
in this task the heavier their masses. This is a widely accepted qualitative
argument and a common sense expectation is that sparticles are lighter than,
for instance, 100 TeV or may be even 10 TeV. 
On the other hand it is very difficult,
if not impossible, to quantify this argument in a convincing and fully 
objective way. One particular measure often used in such discussions
is $\Delta_{a_i}=\frac{a_i}{M_Z}\frac{\partial M_Z}{\partial a_i}$, 
where $a_i$'s are input parameters of the MSSM, but other
measures can be as well considered. In any case it is unclear what is 
quantitatively acceptable as the fine tuning and what is not. 
Moreover, the fine tuning can be discussed only in concrete models for the
soft supersymmetry breaking terms, and any conclusion refers to the particular
model under consideration. It is clear that, in spite of being uncontested
qualitative notion, the naturalness and fine-tuning criteria cannot be used
for setting any absolute upper bounds on the sparticle spectra. Instead,
however, the idea which is promoted in ref.\cite{chank3} is that, any sensible
measure of the amount of fine-tuning becomes an interesting criterion for
at least comparing the relative naturalness of various theoretical models
for the soft mass terms in the MSSM lagrangian, that are consistent with
the stronger and stronger experimental constraints.

In the first place, it has been shown \cite{chank4,chank3,barbieri,strumia} 
that, comparing the situation before
and after LEP, the fine-tuning price in the minimal supergravity model
(that is, with universal soft terms at the GUT scale) has significantly 
increased, largely as a result of the unsuccessful Higgs boson search.
Comparing different values of $\tan\beta$, we find that in this model 
naturalness favours an internediate range. Fine tuning increases for small
values because of the lower limit on the Higgs boson mass and increases
for large values because of the difficulty in assuring correct electroweak
symmetry breaking.
This is shown in Fig.4b. In the intermediate $\tan\beta$ region the fine tuning
price  still remains moderate but would strongly increase with
higher limits on the chargino mass.

In view of the above results and in the spirit of using the fine-tuning
considerations as a message for theory rather than experiment, it is 
interesting to discuss the departures from the minimal supergravity model that
would ease the present (particularly in the low and large $\tan\beta$ region,
where the price is high)
and possibly future fine-tuning problem. The first step in this direction
is to identify the parameters that are really relevant for the Higgs 
potential. It has been emphasized in ref.\cite{dimgiu} that 
scalar masses that enter into the Higgs potential at one-loop level
are only the soft Higgs mass parameters
and the third generation sfermion masses. Thus, breaking the universality
between the first two generation sfermion masses and the remaining scalars,
with the former much heavier, does not cost much of the fine-tuning,
but unfortunately it is not useful now (the present bounds on the first
two generation sfermions are still low enough not to be the source of the fine
tuning in the minimal model ).\footnote{This possibility has been discussed
as a way to ameliorate the FCNC problem in the MSSM.\cite{dimgiu,FCNC} 
One should stress,
however, that it cannot {\it solve} the FCNC problem.}

Furthermore, the attention has recently been drawn \cite{king} 
to the fact that,
at one-loop, the Higgs potential depends  on the gluino mass but not on
the wino and bino masses. This is interesting as it means that in models
with $M_3\neq M_{1,2}$ the fine tuning price is in fact weakly dependent
on the limits on the chargino mass (but not totally independent because of
the constraints on the $\mu$ parameter, which is present in the Higgs potential
at the tree level) and the Tevatron direct bound on the gluino mass is
weaker than the indirect bound obtained from LEP2 assuming gaugino mass
universality. Allowing for
$M_3<M_{1,2}$,  \footnote{As disussed earlier, such a scenario may also be 
interesting for the gauge coupling unification.\cite{roszkowski}}
the after-LEP fine-tuning price is reduced  mainly in the intermediate
$\tan\beta$ region, where it was still quite modest even in the universal
model, but
this possibility may be particularly interesting for intermediate $\tan\beta$
region when the lower limit on the chargino mass is pushed higher.

\begin{figure}
\psfig{figure=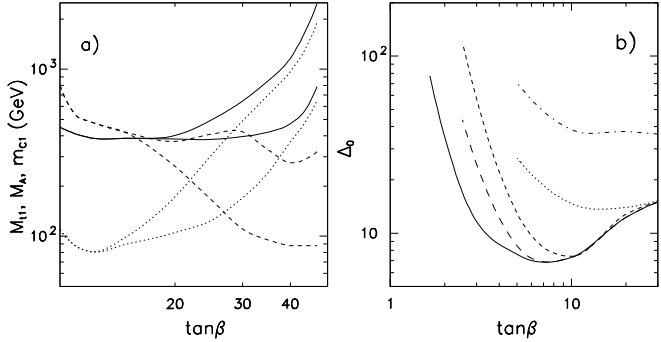,height=5.0cm}
\caption{{\bf a)} Lower   limits on the  lighter    (dotted lines) and
heavier  (solid lines) stop   and on the  $CP-$odd  Higgs  boson $A^0$
(dashed lines)  in the  minimal  supergravity scenario  with  $b-\tau$
Yukawa coupling   unification,   as functions  of   $\tan\beta$. Upper
(lower)  lines   refer to the  case  with   the $b\rightarrow s\gamma$
constraint imposed   (not imposed).  b) Fine-tuning  measures  as 
functions of  $\tan\beta$.
Lower limits  on  the Higgs boson  mass of  90  GeV (solid),   100 GeV
(long-dashed),  105 GeV  (dashed)   110   GeV  (dotted) and   115  GeV
(dot-dashed)  have been assumed. }
\label{fig:barc4}
\end{figure}

After identifying the parameters which are relevant for the Higgs potential
at one-loop level,
that is the Higgs, stop and gluino soft masses and the $\mu$ parameter, 
it is clear that the 
fine tuning price does not increase much even if other superparticles are
substantially heavier. The question remains 
what sort of pattern for soft terms would reduce the fine tuning
caused by the present and future limits on the relevant parameters.
One obvious possibility are non-universal soft Higgs boson masses \cite{marek},
which can significantly  reduce the fine-tuning price 
\cite{chank4,king} and particularly
in the large $\tan\beta$ region \cite{chank3}.
Another alternative is that a model with independent mass 
parameters is in fact simply
inadequate to address the naturalness problem. A relevant example is related
to the so-called $\mu$-problem. One hopes that in the ultimate theory the 
$\mu$ parameter will be calculated  in terms of the soft 
supersymmetry breaking masses. A model which correlates $\mu$ and the gluino
mass may have dramatic effects on the fine tuning price \cite{chank3}, as they 
should not be any more considered as independent parameters. Thus the fine
tuning price will depend on the actual solution to the $\mu$-problem.
\vskip 0.3cm

\section{Conclusions}

Direct searches for superpartners have so far come up empty-handed. 
Nevertheless, we get from experiment a handful of important information
on supersymmetric models, which makes the whole concept of low energy
supersymmetry much more constrained than a decade ago. Simplest ideas
like the minimal supergravity model with universal but otherwise independent
mass terms may soon become inadequate.

For a more complete list of earlier references see, for instance, 
\cite{pokpar,pokwar,KANE1,unpublished}.
\vskip 0.2cm

{\bf Acknowledgments}
This work was supported by Polish State Commitee for Scientific Research under
grant 2 P03B 030 14 (for 1998-99). 
I am greatful to P.H. Chankowski for his help in the preparation of this text.

\end{document}